\documentclass[twocolumn,showpacs,preprintnumbers,amsmath,amssymb, aps,prb,superscriptaddress]{revtex4-1}
\usepackage{graphicx}
\usepackage{dcolumn}
\usepackage{bm}
\begin{document}

\title{Charge transfer in FeOCl intercalation compounds and its pressure dependence: An x-ray spectroscopic study}
\author{I. Jarrige}
\email{jarrige@spring8.or.jp}
\affiliation{Japan Atomic Energy Agency, SPring-8, 1-1-1 Kouto, Sayo, Hyogo 679-5148, Japan}
\author{Y. Q. Cai}
\affiliation{National Synchrotron Light Source II, Brookhaven National Laboratory, Upton, New York 11973-5000, USA}
\author{S. R. Shieh}
\affiliation{Department of Earth Sciences, University of Western Ontario, London, Ontario, N6A 5B7, Canada}
\author{H. Ishii}
\author{N. Hiraoka}
\affiliation{National Synchrotron Radiation Research Center, Hsinchu 30076, Taiwan}
\author{S. Karna}
\author{W.-H. Li}
\affiliation{Department of Physics, National Central University, Jhongli 32001, Taiwan}

\begin{abstract}
We present a study of charge transfer in Na-intercalated FeOCl and polyaniline-intercalated FeOCl using high-resolution x-ray absorption spectroscopy and resonant x-ray emission spectroscopy at the Fe-$K$ edge. By comparing the experimental data with ab-initio simulations, we are able to unambiguously distinguish the spectral changes which appear due to intercalation into those of electronic origin and those of structural origin. For both systems, we find that about 25$\%$ of the Fe sites are reduced to Fe$^{2+}$ via charge transfer between FeOCl and the intercalate. This is about twice as large as the Fe$^{2+}$ fraction reported in studies using M\"{o}ssbauer spectroscopy. This discrepancy is ascribed to the fact that the charge transfer occurs on the same time scale as the M\"{o}ssbauer effect itself. Our result suggests that every intercalated atom or molecule is involved in the charge-transfer process, thus making this process a prerequisite for intercalation. The Fe$^{2+}$ fraction is found to increase with pressure for polyaniline-FeOCl, hinting at an enhancement of the conductivity in the FeOCl intercalation compounds under pressure.
\end{abstract}


\maketitle

\section{Introduction}
Organic electronics holds technological promise for various emerging application areas. This has stimulated a great deal of research into the transport and optical properties of conducting polymers for the past two decades \cite{brutting}. From the characterization viewpoint, the ability of some inorganic layered compounds to intercalate polymers in their intralamellar space is particularly attractive, as it enables confining the polymer chains in a well-defined environment. Besides, the resulting organic-inorganic intercalation hybrids often exhibit novel magnetic and electronic properties.\cite{fujita,sanchez}

At the heart of the interaction between the inorganic host matrix and the intercalated guest lies a charge transfer, which results in partial reduction of the matrix.\cite{schollhorn} A strongly oxidizing host is therefore key to a successful intercalation, providing the driving force for an optimum uptake of guest molecules. As such, iron oxychloride has long been of interest, known to intercalate a great variety of species owing to its high oxidizing power.\cite{bruce} The fact that the charge transfer can induce perturbations in the physical properties of the host underlines the necessity of quantitative measurements of the amount of charge transferred from the guest to FeOCl. Such measurements, using M\"{o}ssbauer spectroscopy, have been previously reported.\cite{fatseas,eckert,herber,philips,kauzlarich,wu} They disclosed that the charge transfer can be understood as a thermally-activated electron hopping process occurring on the M\"{o}ssbauer time scale ($\sim10^{-8}$ s), which can be written as
 \begin{center}
 FeOCl + $n$I $\rightleftharpoons$ Fe$^{3+}_{1-n}$Fe$^{2+}_{n}$ + $n$I$^{+}$
 \end{center}
where I stands for the intercalate, and $n$ is the fraction of the intercalate involved in the charge transfer.

M\"{o}ssbauer spectroscopy has been instrumental in exploring the temperature dependence of the frequency of the hopping process. However, the contributions of the Fe$^{2+}$ and Fe$^{3+}$ fractions to the M\"{o}ssbauer spectrum change as a function of temperature, hindering the quantitative estimation of the $d$ electron count. By employing a faster probe like x-ray absorption spectroscopy, distinct Fe$^{2+}$ and Fe$^{3+}$ states should be observed irrespective of the temperature, thus permitting a quantitative analysis of the charge state free of electronic-relaxation effects. Using x-ray absorption spectroscopy in the partial fluorescence yield mode (PFY-XAS), a probe of the electronic structure with a $\sim10^{-16}$ s time scale, we have quantitatively analyzed the charge transfer in FeOCl intercalated with Na and FeOCl intercalated with polyaniline (designated hereafter as Na-FeOCl and PANI-FeOCl, respectively). Our main finding is that the Fe$^{2+}$ fraction is about 25$\%$ for both intercalation compounds, which is significantly larger than the fractions estimated by M\"{o}ssbauer spectroscopy.\cite{fatseas,eckert,herber,philips,kauzlarich,wu} This result indicates that all the atoms or molecules accommodated in the host lattice are involved in the charge-transfer process. Under external pressure, the degree of charge transfer is observed to increase in PANI-FeOCl, reflecting the increased guest-host interaction.

\section{Experiment}
Polycrystalline iron-oxychloride FeOCl was prepared by the chemical vapor transport technique.\cite{hwang} The composition of the Na-intercalated sample was estimated to be Na$_{0.27}$FeOCl. The composition of the PANI-FeOCl sample used in this study was not measured, but is considered to be very close to the composition of the sample prepared according to the same experimental procedure in Ref. \onlinecite{hwang}, (PANI)$_{0.16}$FeOCl.

The experiment was carried out on the Taiwan beamline BL12XU at SPring-8 in Japan. The undulator beam was monochromatized by a pair of Si(111) mirrors , and focused to a spot size of 120 (horizontal) $\times$ 80 (vertical) $\mu$m$^{2}$ at the sample position using a toroidal mirror. A Rowland-type spectrometer equipped with a 1-m bent Si(531) crystal and a solid-state detector was used to analyze the Fe K$\beta_{1,3}$ ($3p \rightarrow 1s$) emission line. The overall energy resolution was about 1.1~eV. The absorption edges were measured using PFY-XAS, by collecting the intensity at the maximum of the Fe K$\beta_{1,3}$ emission line while scanning the incident energy across the Fe-$K$ edge. The as-obtained spectra benefit from a higher energy resolution than conventional XAS. This technique has proven useful in isolating the $K$ pre-edge of transition metals from the background of the main edge,\cite{rueff1,jarrige1} most notably for metals in an octahedral environment which are characterized by a weak pre-edge feature, such as FeOCl.\cite{choy} Resonant x-ray emission spectroscopy (RXES) data were obtained by measuring the $3p \rightarrow 1s$ radiative decay following an excitation in the vicinity of the $K$ edge.

Full multiple scattering simulations of the corresponding XAS spectra were performed with the FDMNES code \cite{joly} within the muffin-tin approximation. For the high-pressure measurements, we used a Merrill-Bassett type diamond anvil cell, with both the incoming and emitted x-rays passing through the Beryllium gasket. Silicone oil was used as pressure transmitting medium, and pressure was estimated by the ruby fluorescence method.\cite{mao} Since the samples are hydroscopic, all handling was done in a glove-box under Ar atmosphere, and the samples were kept under vacuum throughout the ambient-pressure measurements.

\section{Results and discussion}
\subsection{FeOCl and Na-FeOCl}
\subsubsection{Interpretation of the $K$ edge lineshape}

The PFY-XAS spectra measured at the Fe-$K$ edge for FeOCl and Na-FeOCl are shown in Fig. 1. The main edge, which corresponds to the dipolar transitions $1s \rightarrow 4p$, is presented in the bottom panel. The spectra are normalized in intensity to their area. An enlarged view around the first peak of the derivative of the spectrum, based on which we estimate the energy of the absorption edge, is shown in the top-left panel. Since Fe is in a centrosymmetric environment, the pre-edge arises mainly from $1s \rightarrow 3d$ quadrupolar transitions \cite{arrio}, and is therefore particularly weak ($\sim 1/50$ of the intensity of the main edge). Accordingly, it is plotted separately from the main edge for visibility, in the top-right panel. In the main edge, it is noteworthy that the feature at $\sim$ 7123~eV is well resolved in the present data, whereas it is barely distinguishable from the broad tail of the lower-energy white line in the conventional XAS spectrum in Ref. \onlinecite{choy}. This highlights the usefulness of the resolution gain of the PFY-XAS technique. It is precisely this feature that appears to be the most sensitive to Na intercalation, as it is noticeably stronger in the spectrum of Na-FeOCl. This induces a decrease of the absorption threshold by $\delta E_{1}$=-0.6~eV, as estimated from the peak energies in the derivative spectra shown in the top-left panel. Simultaneously, a slight shift of the maximum of the two white-line features at 7129 and 7134 eV by $\delta E_{2}$=-0.3 eV and $\delta E_{3}$=-0.5~eV, respectively, is caused by Na intercalation.
\begin{figure}
\includegraphics [width=8.5 cm]{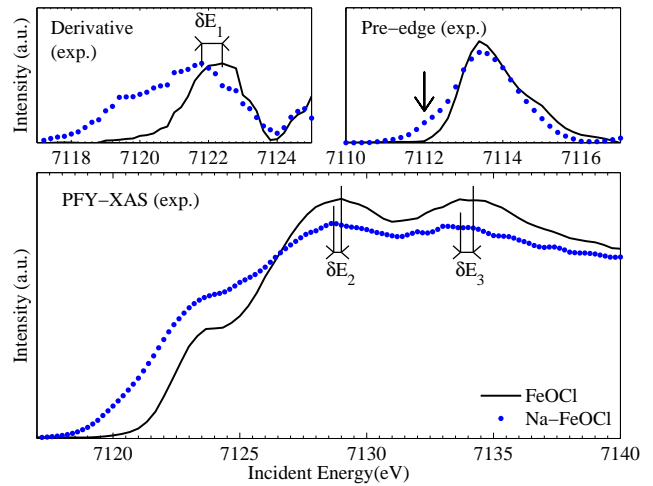}
\caption{(Color online) (bottom) PFY-XAS spectrum measured for FeOCl (solid line) and Na-FeOCl (dots). (top left) Enlarged view of the low-energy peak of the first derivative of the PFY-XAS spectra. (top right) Enlarged view of the pre-edge region. The arrow points to the low-energy shoulder which appears on substitution, attributed to the $^{4}$T$_{1g}$ state of high-spin Fe$^{2+}$.}
\end{figure}

The $K$ edge of transition metals is known to contain both structural and electronic information, sometimes rendering its interpretation difficult.\cite{monesi} This is especially the case here since the intercalation process in FeOCl involves both structural and electronic changes. In order to help discern between the two, we performed numerical simulations of the Fe-$K$ edge using the FDMNES code. With this program, the density of states are calculated within the full multiple scattering framework, and the transition matrix elements are computed in the dipole approximation for the absorption process between the initial state $1s^{2}4p^{0}$ and the final state $1s^{1}4p^{1}$. All simulations were performed for a cluster radius of 4.0$\AA$. The spectrum simulated for FeOCl using the lattice parameters reported in Ref. \onlinecite{hwang}, $a$=3.29921$\AA$, $b$=3.77450$\AA$, $c$=7.90852$\AA$, is shown in Fig. 2. The relative energy positions and intensities of the main spectral features are seen to be in good agreement with the experimental spectrum.

\begin{figure}
\includegraphics [width=8.5 cm]{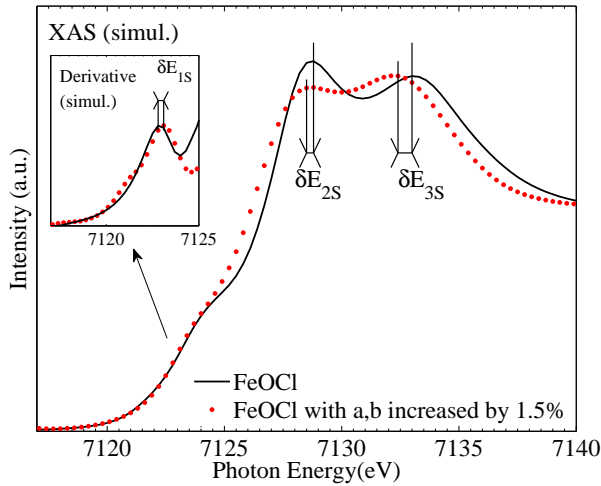}
\caption{(Color online) (top) FDMNES simulations of the XAS spectrum for FeOCl using the structure of the pristine (solid line) and intercalated (dots) materials. An enlarged view of the low-energy peak of the first derivative of the simulated spectra is shown in the inset.}
\end{figure}

The intercalation process has been commonly reported to cause a slight expansion of the lattice parameters $a$ and $b$, while $c$ is doubled and expanded.\cite{eckert,philips,hwang} The doubling of $c$ is ascribed to a shift of ($a$/2,$b$/2) in the stacking of the bilayers, while the expansion, varying between 30 and 70 $\%$ depending on the studies, is that of the intercalation gap. Since the interlayer interactions are not included in our simulations, we used the same value of the $c$ parameter for the spectrum of pristine and intercalated FeOCl. An expansion of $a$ and $b$ by 1.5$\%$ with respect to FeOCl was considered. This is the expansion reported for PANI-FeOCl in Ref. \onlinecite{hwang} and is in line with the reports for other intercalates.\cite{eckert,philips,yao} In Fig. 2, the comparison between the spectra simulated for the pristine and expanded lattices enables us to pinpoint the intercalation-induced changes in the $K$-edge spectrum that result solely from structural changes, without any modification of the Fe $d$ electron count.

The simulated spectra are found to emulate remarkably well the $\delta E_{2}$ and $\delta E_{3}$ shifts of the white lines towards lower energies observed between the experimental spectra of FeOCl and Na-FeOCl (cf. $\delta E_{2}$ and $\delta E_{3}$ in Fig. 1 and $\delta E_{2S}$ and $\delta E_{3S}$ in Fig. 2). Those shifts are plainly consistent with the fact that the longer the bond the lower the bond energy. The 7123-eV structure, on the other hand, seems to be nearly unaffected by the lattice expansion. The low-energy peak of the first derivative is found to shift slightly towards higher energies, probably because of the increased overlap with the tail of the 7129-eV peak. This structural shift is estimated to be $\delta E_{1S}$=+0.3~eV (cf. inset of Fig. 2). This demonstrates that the low-energy shift $\delta E_{1}$=-0.6~eV observed between the edges of the experimental spectra of FeOCl and Na-FeOCl is purely of electronic origin, corresponding to the partial reduction of the Fe sites upon intercalation. We deduce this electronic shift of the edge to be about $\delta E_{1E}$=$\delta E_{1}$-$\delta E_{1S}$=-0.9~eV.
\\
\subsubsection{Degree of charge transfer: Estimation}

Having identified the salient changes in the absorption spectrum due to Na intercalation, we now turn to the semiquantitative analysis of the charge transfer between Na and the host matrix. We first limit our attention to the main edge. It is widely accepted that a linear relation between the $K$ edge shift and the valence state can be drawn for metals that have the same ligand.\cite{wong,sakurai} The edge shift was reported to be 4.5~eV between Fe$^{2+}$ and Fe$^{3+}$ for Fe atoms coordinated to oxygen ligands, and 3.7~eV for chlorine ligands \cite{sakurai}. Assuming that in the oxychloride case the edge shift is intermediate between these two values ($\sim$4.1~eV), we estimate from the edge shift $\delta E_{1E}$=0.9~eV that about 22\% of the Fe sites are reduced to Fe$^{2+}$ in Na-FeOCl.

The good agreement between the experiment and the simulations for the structural dependence of the absorption lineshape motivated us to extend the FDMNES calculations to the study of the charge transfer. To this end, we performed calculations of the Fe-$K$ edge for an artificial compound with the composition FeCl$_{2}$ isostructural to intercalated FeOCl, in order to mimic the spectral changes associated with the increase in the Fe $d$ count upon intercalation. This compound is named hereafter FeClCl. A comparison of the spectra simulated for FeClCl and FeOCl is shown in Fig. 3. The spectrum of FeClCl exhibits a notable shift of spectral weight towards lower energies, with the white line pointing at 7127~eV and a low-energy shoulder located around 7122~eV. We point out the resemblance between the simulated edges of FeOCl and FeClCl and the spectra measured for andradite and siderite in Ref. \onlinecite{rueff1}, respectively. Andradite and siderite are minerals which contain high-spin Fe in an octahedral environment, trivalent for the former and divalent for the latter. These similarities lend experimental support to our simulations. We find that a linear combination of the simulated spectra 0.72$\times$intercalated-FeOCl+0.28$\times$FeClCl yields an absorption edge lower than that of simulated FeOCl by 0.6~eV (cf. top-left panel of Fig. 3), equal to $\delta E_{1}$. This suggests a Fe$^{2+}$ fraction of about 28\% in Na-FeOCl, in good agreement with the estimate based on the edge shift presented above.

\begin{figure}
\includegraphics [width=8.5 cm]{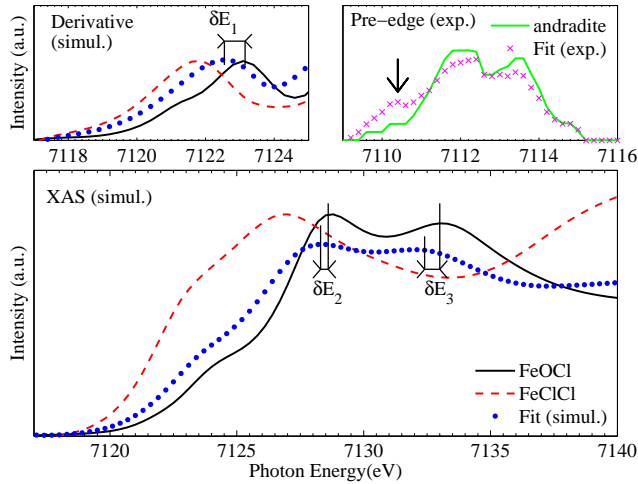}
\caption{(Color online) (bottom) FDMNES simulations of the XAS spectrum for pristine FeOCl (solid line) and the artificial compound FeClCl (dashed line), along with the linear combination of simulated spectra 0.72$\times$intercalated-FeOCl+0.28$\times$FeClCl (dots). (top left) Enlarged view of the low-energy peak of the first derivative of the spectra shown in the bottom panel. (top right) Enlarged view of the pre-edge region of the PFY-XAS spectrum measured for andradite, and the linear combination 0.73$\times$andradite+0.27$\times$siderite. The andradite and siderite spectra are taken from Ref. \onlinecite{rueff1}. The arrow indicates the low-energy shoulder ascribed to the $^{4}$T$_{1g}$ state of high-spin Fe$^{2+}$.}
\end{figure}

To complete our analysis of the charge transfer we now take a moment to disuss the pre-edge. The average centroid of the pre-edge structure has been shown to be correlated in a linear fashion with the oxidation state of the metal provided the coordination number remains unchanged.\cite{wilke,petit} As seen in the top-right panel of Fig. 1, the pre-edge feature exhibits a shift of spectral weight towards lower energies between FeOCl and Na-FeOCl, indicated by the arrow. We attribute this to the growth of the $^{4}$T$_{1g}$ state of high-spin Fe$^{2+}$.\cite{westre} We estimate the associated shift of the centroid to be -0.3~eV. Based on a previous estimation of the centroid shift between Fe$^{2+}$ and Fe$^{3+}$ of 1.4~eV,\cite{wilke,petit,farges,heijboer} we estimate that the -0.3-eV shift corresponds to a reduction of 21\% of the Fe sites due to Na intercalation. As an alternative approach, we also used a linear combination of the pre-edges of andradite and siderite from Ref. \onlinecite{rueff1}. A centroid shift of -0.3~eV is obtained for the linear combination 0.73$\times$andradite (Fe$^{3+}$)+0.27$\times$siderite (Fe$^{2+}$), whose spectrum is shown in the top-right panel of Fig. 3.

\subsubsection{Degree of charge transfer: Discussion}

The values of the Fe$^{2+}$ fraction in Na-FeOCl estimated by the different methods presented in the previous subsection are summarized in Table 1. The average fraction is $\sim$0.25. We estimate the error on each of these individual fractions to be in the order of $\pm$ 0.05, and propose to use the range of the different estimates, $\pm$ 0.035, as the error on the average fraction. Our value of the Fe$^{2+}$ fraction is about twice as large as the estimates based on M\"{o}ssbauer spectroscopy for different guests and various intercalation stoichiometries between 0.05 and 0.36.\cite{fatseas,eckert,herber,philips,kauzlarich,wu} It is well known that the host-matrix charge transfer corresponds to an electron hopping mechanism with a time scale on par with that of the M\"{o}ssbauer effect ($\sim10^{-8}$ s).\cite{fatseas,eckert,herber} Since this is a thermally-activated process, the fraction of Fe$^{2+}$ and Fe$^{3+}$ sites that are undistinguishable and appear as a mixed-valence component in the M\"{o}ssbauer spectrum increases with temperature. Accordingly, the previously reported Fe$^{2+}$ fractions were estimated using the spectrum collected at the lowest possible temperature, just before the onset of antiferromagnetism near $T_{N}$=65 K.\cite{fatseas,eckert,herber,philips,kauzlarich,wu} Below this temperature, the magnetic order splits the spectral lines into unresolved sextets, thereby precluding any quantitative determination of the Fe$^{2+}$/Fe$^{3+}$ ratio. It is however likely that the spectra collected just above $T_{N}$ are still affected by the electron-hopping relaxation time, resulting in lower Fe$^{2+}$ fractions.\cite{fatseas,herber} Following this idea, we extrapolated the temperature dependence of the Fe$^{2+}$ fraction reported by Fatseas \textit{et al.}\cite{fatseas} towards lower temperatures, and obtained a fraction of 0.25 at 0 K, in perfect agreement with our estimation. While reconciling the M\"{o}ssbauer and PFY-XAS analyses, this observation underscores the importance to resort to x-ray spectroscopic probes for quantitative studies of thermal fluctuating valence states. We note that this work constitutes, to our knowledge, the first quantitative estimation of the charge transfer in an FeOCl intercalate using an x-ray spectroscopic technique.

\begin{table}
\caption{Fraction of Fe$^{2+}$ estimated in this study by different methods: Shift of the pre-edge centroid (pre-edge shift), linear combination of the pre-edge spectra of andradite and siderite (pre-edge fit), shift of the low-energy peak of the first derivative of the absorption spectrum (main edge shift), and linear combination of the simulated absorption edges (main edge fit).}
\begin{center}
\begin{ruledtabular}
  \begin{tabular}{ccccc}
    &\multicolumn{2}{c}{pre-edge}&\multicolumn{2}{c}{main edge}\\
    \raisebox{1.5 ex} {Method} &shift&fit&shift&fit \\ [0.5 ex]
    \hline
    Fraction of Fe$^{2+}$& 0.21 & 0.27 & 0.22 & 0.28 \\
  \end{tabular}
\end{ruledtabular}
\end{center}
\end{table}

Our estimate of the Fe$^{2+}$ fraction in Na$_{0.27}$FeOCl interestingly suggests that the ionization of the guest material is complete within the analysis error. This is in contrast with the long-held view based on M\"{o}ssbauer results that the ionisation is partial, irrespective of the nature of the guest and of the intercalation stoichiometry.\cite{eckert,herber,kauzlarich} Whereas those studies implied that the charge transfer process is limited by the relatively small number of electrons that the FeOCl layers can accept, our study rather suggests that the charge transfer, because involving each intercalated Na atom, acts as a prerequisite for intercalation. While this scenario has yet to be confirmed by a systematic measurement of FeOCl with different guest species and stoichiometries, it is qualitatively supported by the fact that FeOCl has about the highest oxidizing power among the generally used layered host lattices. \cite{bruce} It is then conceivable that other less reactive host compounds, on the other hand, only partially reduce the guest material.

\subsection{FeOCl and PANI-FeOCl}
\subsubsection{Ambient pressure}

The $1s3p$-RXES spectra measured across the $K$-edge for FeOCl and PANI-FeOCl are presented in Fig. 4 along with the absorption spectra measured in the PFY mode. The RXES spectra are plotted as a function of the transfer energy, which is defined as the difference between the incident and emitted photon energies. They are vertically shifted in correspondence with the energy axis of the PFY-XAS spectrum. The energy of the edge of PANI-FeOCl is found to coincide with that of Na-FeOCl, which suggests a Fe$^{2+}$ fraction of $\sim$0.25 in PANI-FeOCl too. This fraction is greater than the intercalation stoichiometry, $\sim$0.16, which can be explained by the presence of a lone pair of electrons in PANI, possibly resulting in a stronger donor ability than Na. Besides, this result agrees with our conjecture that every intercalated molecule is involved in the charge transfer.

\begin{figure}
\includegraphics [width=8.5 cm]{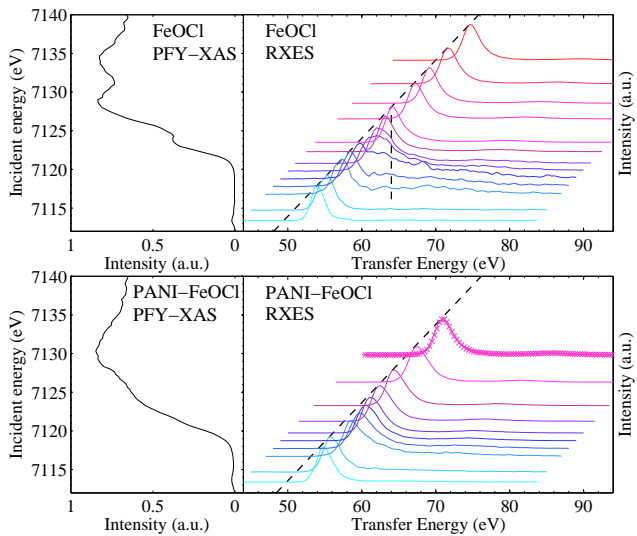}
\caption{(Color online) Incident-energy dependence of the $1s3p$-RXES spectra obtained for FeOCl (top) and PANI-FeOCl (bottom), along with their respective PFY-XAS spectrum. The vertical offset of the RXES spectra corresponds to the incident-energy axis of the PFY-XAS spectra. The oblique and vertical dashes lines highlight the fluorescent and Raman-like dispersions, respectively. The RXES spectrum measured for FeOCl at 7130~eV is superimposed to that of PANI-FeOCl in the bottom panel (crosses).}
\end{figure}

$K$-edge RXES can been used as a tool to qualitatively characterize the degree of localization of the empty $3d$ and $4p$ states of transition metals. \cite{rueff1,chen,yamaoka} This is done by separating the Raman and fluorescence channels which coexist in the RXES spectra. These two channels exhibit distinct behaviors upon increasing the incident energy, as the Raman signal stays at constant transfer energy while the fluorescence disperses linearly towards higher transfer energies. The Raman process is considered to be a coherent second-order optical process, corresponding to transitions to a localized intermediate state $1s^{1}3d^{n+1}$. In contrast, the non-coherent fluorescence is usually ascribed to transitions to a delocalized intermediate state $1s^{1}3d^{n}4p^{n+1}$. The incident energy of the crossover between the Raman and the fluorescence regimes yields information about the relative extents of the localized and band states. For both FeOCl and PANI-FeOCl, the transfer energy of the main peak is found to track the incident energy throughout the range of incident energies of the experiment, as indicated by the oblique dashed lines in Fig. 4. The absence of Raman behavior for the main peak down to the pre-edge region points to delocalized unoccupied Fe $3d$ states. This delocalized character can be readily explained by strong hybridization between the Fe $3d$ states and the $2p$ states of the neighboring O and Cl atoms. Furthermore it shows that the electronic structure of FeOCl, semiconducting with a 1.9-eV gap,\cite{kanamaru,kauzlarich} is not governed by strong electronic correlations.

In the spectra of FeOCl, we discern some additional spectral weight emerging around a fixed transfer energy of $\sim$64~eV between the pre-edge and the main edge. Its intensity increases with the incident energy, until it merges with the main, fluorescent-like signal around 7123~eV. Since this incident energy corresponds to the lowest-energy feature of the main edge, we interpret this Raman signal in terms of transitions into the lowest-unoccupied, narrow Fe $4p$ band. Hence, from the perspective of RXES, the low-energy unoccupied Fe $4p$ states of FeOCl appear to retain a somewhat localized character. We notice that this Raman feature has virtually disappeared in the PANI-intercalated compound, suggesting that the lowest-unoccupied $4p$ band is more diffuse than in FeOCl. While the $3d$ states are known to be the electron acceptor level\cite{kim} associated with the increased conductivity on intercalation,\cite{kauzlarich,wu} this finding interestingly reveals that the enhanced electronic itinerancy in the intercalation compound is reflected in the $4p$ band too. Finally, the spectrum taken on FeOCl at an incident energy of 7130~eV is superimposed with that of PANI-FeOCl in the lower panel of Fig. 4. Both lineshapes are found to be identical, and the relative intensity of the K$\beta$' satellite around 86-eV seems consistent with a high-spin state.\cite{rueff3}

\subsubsection{High pressure}

Optimizing the electronic conductivity of the FeOCl intercalation compounds is essential for their potential application as cathode material in secondary batteries.\cite{sagua} PANI intercalation results in a $p$-type semiconductor behavior and a substantial increase in the conductivity in the order of 10$^{5}\sim$ 10$^{6}$ $\Omega^{-1}$.cm$^{-1}$.\cite{wu,scully} The fact that the conductivity of the intercalated FeOCl systems is governed by the $3d$ electron hopping mechanism in the partially reduced FeOCl layers \cite{bruce} suggests that increasing the extent of charge transfer between the guest and the host could enhance the conductivity. One may instinctively expect the charge transfer to increase with increasing the host-guest interaction. To test this idea, we have measured the Fe-$K$ PFY-XAS spectrum on PANI-FeOCl subjected to external pressure.

\begin{figure}
\includegraphics [width=8.5 cm]{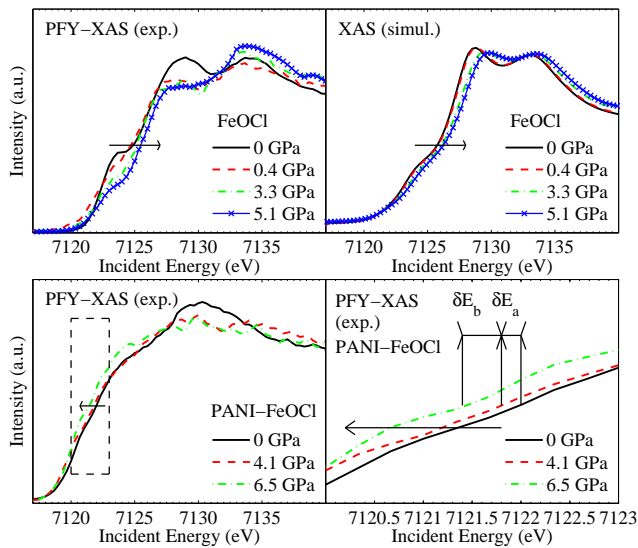}
\caption{(Color online) (top left) Pressure dependence of the PFY-XAS experimental spectrum of FeOCl. (top right) Simulation of the pressure dependence of the XAS spectrum of FeOCl using FDMNES. (bottom left) Pressure dependence of the PFY-XAS experimental spectrum of PANI-FeOCl. (bottom right) Enlarged view of the region delineated by the dashed lines in the bottom left panel. In all plots, the arrow indicates the direction of increasing pressure.}
\end{figure}

We start by presenting the high-pressure data obtained on pristine FeOCl. The pressure dependence of both experimental and simulated Fe-$K$ edges is presented in the top panels of Fig. 5. The spectra are normalized in intensity to their area. For the simulations, since no high-pressure structural data on FeOCl is available in the literature, we used the relative changes in the lattice parameters of TiOCl under pressure reported in Ref. \onlinecite{forthaus}. The compression was found to be anisotropic, largest along $c$. In the experimental PFY-XAS spectra, as pressure is increased up to 5.1~GPa, an edge shift is seen to occur as a result of transfer of spectral weight from the two lower-energy structures towards the 7135-eV centered peak. Because lower-energy features correspond to longer metal-ligand bonds, the growth of the 7135-eV peak at the expense of the two lower-energy peaks indicates that the longer basal Fe-O and Fe-Cl bonds are more affected by the compression than the apical, shorter Fe-O bonds parallel to the $a$ axis. Although the simulated spectra reproduce qualitatively the pressure-induced edge shift, the 7135-eV peak is found to shift too and its intensity is less affected than in the experiment. The discrepancy with the experiment could be indicative of an even more anisotropic compression in FeOCl than in TiOCl.

The pressure dependence of the experimental spectrum of the PANI-intercalated compound shows an opposite behavior, with a decrease of the edge energy under pressure, as seen in the bottom panels of Fig. 5. The comparison with FeOCl points to an electronic origin of the edge shift for PANI-FeOCl, i.e., an increase of the ratio of the reduced Fe$^{2+}$ sites as pressure is increased. The peak in the first derivative did not show a clear change as a function of pressure. Alternatively, we tentatively used the shift of the edge at half-maximum ($\delta E_{a}$ and $\delta E_{b}$ in Fig. 5) to make a rough estimate of the Fe$^{2+}$ fraction, which we find to increase from 0.25 at ambient pressure to $\sim$ 0.3 at 4.1 GPa and $\sim$ 0.4 at 6.5 GPa. Although these values are approximate and probably overestimated, they seem to suggest that all the PANI molecules have donated their two electrons to the FeOCl matrix at 6.5 GPa.

Such a pressure-enhanced charge transfer is in line with the intuitive idea that, as the contact area between PANI and FeOCl increases under compression, electrons are being drawn from PANI to FeOCl with more ease. Besides, the electrons donated by the guest were proposed to go in an unoccupied antibonding Fe-O orbital using cluster calculations.\cite{kim} Under pressure the crystal field splitting of the Fe site increases, which should cause the lowest unoccupied minority-spin $t_{2g}$ level, and consequently the acceptor Fe-O level, to be shifted down in energy. This process too should favor the charge transfer with PANI. We note that a previous study on the graphite intercalation compound C$_{24}$SbCl$_{5}$ reported an increased conductivity both along the $c$ axis and in the basal plane under pressure.\cite{iye} Whereas the increase of the $c$-axis conductivity was ascribed to the decrease in the interlayer separation, no sound explanation was proposed for the increase of the basal-plane conductivity. Based on our results, it would seem plausible that an increase in the degree of charge transfer accounts for the increase of the basal-plane conductivity in the graphite intercalation compound.

\section{Conclusion}
We have estimated the degree of charge transfer in Na-FeOCl and PANI-FeOCl using PFY-XAS at the Fe-$K$ edge. For both intercalates, we find that the fraction of Fe sites reduced to Fe$^{2+}$ amounts to about 25\%. This result contrasts with previous studies using M\"{o}ssbauer spectroscopy which reported substantially smaller Fe$^{2+}$ fractions. Because the charge transfer and the M\"{o}ssbauer effect have comparable time characteristics, we believe that our estimation, employing a significantly faster probe, is more reliable. Our estimated Fe$^{2+}$ fraction is on par with (Na) or larger than (PANI) the intercalate concentration. This shows that, contrary to what has been thought so far, the charge transfer involves every atom or molecule of the intercalate, therefore stressing the crucial role played by the charge transfer in the intercalation process. Using RXES, we have observed signatures of electron delocalization in the Fe unoccupied $4p$ states upon PANI intercalation. Finally, the last part of our study disclosed that the charge transfer in PANI-FeOCl can be enhanced through the application of external pressure. This finding is evocative of a pressure-induced increase in conductivity in the FeOCl intercalated compounds, and could therefore be of interest for their application as cathode material.

\begin{acknowledgments}
We thank C.C. Chen of NSRRC for his technical support, and S. Tsutsui and P. Kr\"{u}ger for helpful discussions about M\"{o}ssbauer spectroscopy and FDMNES simulations, respectively. This work was performed with the approvals of JASRI/SPring-8 (Proposal No. 2005B4261) and NSRRC, Taiwan (2004-3-074-1), and is partly supported by NSRRC, the National Science Council of Taiwan (NSC 94-2112-M-213-012).
\end{acknowledgments}


\begin{thebibliography}{00}
\bibitem{brutting} \textit{Physics of Organic Semiconductors}, edited by W. Br\"{u}tting (Wiley-VCH, Berlin, 2005).
\bibitem{fujita} W. Fujita and K. Awaga, J. Am. Chem. Soc. {\bf119}, 4563 (1997).
\bibitem{sanchez} V. S\'{a}nchez, E. Benavente, M.A. Santa Ana, and G. Gonz\'{a}lez, Chem. Mater. {\bf11}, 2296 (1999).
\bibitem{schollhorn} R. Sch\"{o}llhorn, Angew. Chem. {\bf19}, 983 (1980).
\bibitem{bruce} D. O'Hare in \textit{Inorganic Materials}, edited by D. W. Bruce and D. O'Hare (John Wiley $\&$ Sons Ltd., Chichester, 1997), Chap. 4, pp. 202-208.
\bibitem{fatseas} G.A. Fatseas, P. Palvadeau, and J.P. Venien, J. Sol. Stat. Chem., {\bf51}, 17 (1984).
\bibitem{eckert} H. Eckert, R.H. Herber, J. Chem. Phys. {\bf80}, 4526 (1984).
\bibitem{herber} R.H. Herber, H. Eckert, Phys. Rev. B {\bf31}, 34 (1985).
\bibitem{philips} J.E. Philips and R.H. Herber, Inorg. Chem. {\bf25}, 3081 (1986).
\bibitem{kauzlarich} S.M. Kauzlarich, J.F. Ellena, P.D. Stupik, W.M. Reiff, and B.A. Averill, J. Am. Chem. Soc. {\bf109}, 4561 (1987).
\bibitem{wu} C.-G. Wu, D.C. DeGroot, H.O. Marcy, J.L. Schindler, C.R. Kannewurf, T. Bakas, V. Papaefthymiou, W. Hirpo, J.P. Yesinowski, Y.-J. Liu, and M.G. Kanatzidis, J. Am. Chem. Soc., {\bf117}, 9229 (1995).
\bibitem{hwang} S.R. Hwang, W.-H. Li, K.C. Lee, J.W. Lynn, C.-G. Wu, Phys. Rev. B {\bf62}, 14157 (2000).
\bibitem{jarrige1} I. Jarrige, Y.Q. Cai, H. Ishii, N. Hiraoka, A. Bleuzen, Appl. Phys. Lett. {\bf93}, 054101 (2008).	
\bibitem{rueff1} J.-P. Rueff, L. Journel, P.-E. Petit, F. Farges, Phys. Rev. B {\bf69}, 235107 (2004).
\bibitem{choy} J.-H. Choy, J.-B. Yoon,  D.-K. Kim, S.-H. Hwang, Inorg. Chem. {\bf34}, 6524 (1995).
\bibitem{joly} Y. Joly, Phys. Rev. B {\bf63}, 125120 (2001).
\bibitem{mao} H.K. Mao, P.M. Bell, J.W. Shaner, and D.J. Steinberg, J. Appl. Phys. {\bf49}, 3276 (1978).
\bibitem{arrio} M.-A. Arrio, S. Rossano, Ch. Brouder, L. Galoisy, and G. Calas, Europhys. Lett. {\bf51}, 454 (2000).
\bibitem{monesi} C. Monesi, C. Meneghini, F. Bardelli, M. Benfatto, S. Mobilio, U. Manju, D.D. Sarma, Nucl. Instr. and Meth. in Phys. Res. B {\bf246}, 158 (2006).
\bibitem{yao} Y.-D. Dai, W. Han, L. Zheng, Y.-F. Xia, Chin. Phys. Lett. {\bf22}, 2059 (2005).
\bibitem{wong} J. Wong, F.W. Lytle, R.P. Messmer, and D.H. Maylotte, Phys. Rev. B {\bf30}, 5596 (1984).
\bibitem{sakurai} K. Sakurai, A. Iida, and Y. Gohshi, Anal. Sci. {\bf4}, 37 (1988).
\bibitem{wilke} M. Wilke, F. Farges, P.E. Petit, G.E. Brown Jr., and F. Martin, Am. Mineral. {\bf65}, 713 (2001).
\bibitem{petit}P.-E. Petit, F. Farges, M. Wilke, and V.A. Sol\'e, J. Synch. Rad. {\bf8}, 952 (2001).
\bibitem{westre} T.E. Westre, P. Kennepohl, J.G. DeWitt, B. Hedman, K.O. Hodgson, and E.I Solomon, J. Am. Chem. Soc., {\bf119}, 6297 (1997).
\bibitem{farges} F. Farges, Y. Lefr\`ere, S. Rossano, A. Berthereau, G. Calas, and G.-E. Brown Jr., J. Non-cryst. Sol. {\bf344}, 176 (2004).
\bibitem{heijboer} W.M. Heijboer, P. Glatzel, K.R. Sawant, R.F. Lobo, U. Bergmann, R.A. Barrea, D.C. Koningsberger, B.M. Weckhuysen, and F.M.F. de Groot, J. Phys. Chem. B. {\bf108}, 10002 (2004).
\bibitem{chen} J.M. Chen, C.K. Chen, T.L. Chou, I. Jarrige, H. Ishii, K.T. Lu, Y.Q. Cai, K.S. Liang, J.M. Lee, S.W. Huang, T.J. Yang, C.C. Shen, R.S. Liu, J.Y. Lin, H.T. Jeng, and C.-C. Kao, Appl. Phys. Lett. {\bf91}, 054108 (2007).
\bibitem{yamaoka} H. Yamaoka, N. Tsujii, H. Oohashi, D. Nomoto, I. Jarrige, K. Takahiro, K. Ozaki, K. Kawatsura, and Y. Takahashi, Phys. Rev. B {\bf77}, 115201 (2008).
\bibitem{kanamaru} F. Kanamaru, M. Shimada, M. Koizumi, M. Takano, and T. Takada, J. Solid State Chem. {\bf7}, 297 (1973).
\bibitem{kim} S.-H. Kim and H. Kim, Bull. Korean Chem. Soc. {\bf14}, 132 (1993).
\bibitem{rueff3} J.-P. Rueff, A. Shukla, A. Kaprolat, M. Krisch, M. Lorenzen, F. Sette, R. Verbeni, Phys. Rev. B {\bf63}, 132409 (2001).
\bibitem{sagua} A. Sagua, E. Moran, M.A. Alario-Franco, A. Rivera, C. Leon, J. Santamaria, and J. Sanz, Int. J. Inorg. Mater. {\bf3}, 293 (2001).
\bibitem{scully} S.F. Scully, R. Bissessur, D.C. Dahn, and G. Xie, Sol. Stat. Ion. {\bf181}, 933 (2010).
\bibitem{forthaus} M.K. Forthaus, T. Taetz, A. M\"{o}ller, and M.M. Abd-Elmeguid, Phys. Rev. B {\bf77}, 165121 (2008).
\bibitem{iye} Y. Iye, O. Takahashi, S. Tanuma, K. Tsuji, and S. Minomura, J. Phys. Soc. Jpn. {\bf51}, 475 (1982).
\end{thebibliography}
\end{document}